\documentclass[showpacs,aps,twocolumn]{revtex4}
\usepackage{epsfig}
\usepackage{graphicx}
\usepackage{amsmath,amssymb,amsfonts}
\usepackage{array}
\usepackage{url}
\usepackage[dvipsnames]{xcolor}
\usepackage{multirow}
\usepackage{float}
\usepackage{subcaption}
\usepackage{lineno}
\usepackage{xspace}
\usepackage{ulem}

\newcommand{\bef}{\begin{figure}}
\newcommand{\eef}{\end{figure}}
\newcommand{\bc}{\begin{center}}
\newcommand{\ec}{\end{center}}
\newcommand{\nn}{\nonumber}
\newcommand{\be}{\begin{equation}}
\newcommand{\ee}{\end{equation}}
\newcommand{\bea}{\begin{eqnarray}}
\newcommand{\eea}{\end{eqnarray}}

\def\ba{\begin{eqnarray}}
\def\ea{\end{eqnarray}}

\definecolor{darkblue}{RGB}{0,0,196}
\usepackage[colorlinks=true,linkcolor=darkblue,citecolor=darkblue,urlcolor=darkblue]{hyperref}

\begin{document}
\title{Development of next-generation light-weight ternary Mg--Al--Li alloys for beampipe applications in particle accelerators}
\author{Kamaljeet Singh$^{1}$}
\author{Kangkan Goswami$^{1}$}
\author{Raghunath Sahoo$^{1}$}
\email[Corresponding Authors: ]{Raghunath.Sahoo@cern.ch}
\author{Sumanta Samal$^{2}$}
\email[]{sumanta@iiti.ac.in}

\affiliation{$^{1}$Department of Physics, Indian Institute of Technology Indore, Simrol, Indore 453552, India}
\affiliation{$^{2}$Department of Metallurgical Engineering and Materials Science, Indian Institute of Technology Indore, Simrol, Indore 453552, India}

\begin{abstract}
The current study reports the design of advanced light-weight materials for high-energy accelerator beampipe applications. The objective is to optimize the combined requirements of high radiation length and stiffness properties of the designed materials. The present study targets conventional beampipe materials such as aluminum, titanium, and stainless steel as primary performance benchmarks. These conventional beampipes are used at synchrotron radiation sources, such as Indus-1 and Indus-2 in India, the Nuclotron-based Ion Collider Facility in Russia, and the ring synchrotron facility SIS 100/300 at the Facility for Antiproton and Ion Research in Germany.
In this context, a series of ternary Mg--Al--Li alloys is systematically investigated to enhance the figure of merit.
Two aluminum-rich alloys, A1 ($\mathrm{Al_{61.5}Li_{10.8}Mg_{27.7}}$) and A2 ($\mathrm{Al_{66}Li_{19.4}Mg_{14.6}}$), along with three magnesium-rich alloys, M1 ($\mathrm{Al_{23.9}Li_{29.3}Mg_{46.8}}$), M2 ($\mathrm{Al_{19}Li_{20.6}Mg_{60.4}}$), and M3 ($\mathrm{Al_{39.8}Li_{20.1}Mg_{40.1}}$) are explored.
Thermodynamic stability, density, liquidus temperature, and phases are evaluated using Latin hypercube sampling within the Thermo-Calc TC-Python framework. Elastic properties are obtained from density functional theory calculations performed using the Vienna \textit{Ab Initio} Simulation Package. Our results show that, although the elastic moduli ($E$) of the investigated Mg-Al–Li alloys are comparable to those of conventional beampipe materials, their significantly higher radiation lengths ($X_0$) lead to an overall improvement in the figure of merit $X_0 E^{1/3}$.
These findings demonstrate that Mg--Al--Li alloys constitute a promising material class for beampipe applications in particle accelerators. These alloys offer an improved balance between radiation transparency and mechanical performance relative to conventional materials.
\end{abstract}

\date{\today}
\maketitle

\section{Introduction}
Particle accelerators play a crucial role in advancing fundamental particle physics, enabling the exploration of subatomic particles and the fundamental forces governing the universe. Modern accelerators span a wide range of scales, from compact table-top devices such as medical linear accelerators used for cancer radiotherapy and ion accelerators employed in material synthesis and characterization, to large-scale research facilities like the Large Hadron Collider (LHC) and the Relativistic Heavy Ion Collider (RHIC)~\cite{Sahoo:2021aoy, Busza:2018rrf}. At the heart of every accelerator, there is a carefully engineered cylindrical structure component, known as the beampipe that guides accelerated particles along their intended trajectories. The beampipe must sustain an ultra-high vacuum environment to ensure that particles propagate over long distances without undergoing unwanted collisions with residual gas molecules, which would otherwise lead to beam scattering, energy loss, and production of secondary radiation~\cite{art, Burkhardt2012}.

An ideal beampipe material for high-energy collider experiments is required to possess a large radiation length ($X_0$), sufficient mechanical robustness, and compatibility with ultra-high vacuum conditions. The radiation length quantifies the transparency of a material to electromagnetic radiation and is defined as the mean distance over which a high-energy electron loses $1/e$ of its energy via bremsstrahlung. Since $X_0$ scales approximately as $1/Z^{2}$, where $Z$ is the atomic number, low-$Z$ materials are naturally favored to minimize multiple scattering and energy degradation of the beam~\cite{DaSilva2025, knaster, CORREA2018291}. 
In addition to radiative considerations, the elastic modulus ($E$) plays a decisive role in ensuring the structural integrity of the beampipe under pressure gradients, thermal loads, and mechanical constraints, particularly near the interaction region where thin walls and small diameters are mandatory~\cite{Veness:2011zz,bruning2015high}. These effects are critical near the interaction region, where even small amounts of material can significantly degrade tracking resolution and vertex reconstruction accuracy. Quantitatively, this trade-off can be captured using the figure of merit $X_{0}E^{1/3}$, which combines radiation transparency with mechanical stiffness.

As a perfect element, lithium represents the lowest-$Z$ metallic element, which makes it an attractive candidate from the standpoint of radiation length. However, pure lithium is impractical for accelerator applications due to its extremely low melting point, high chemical reactivity, and insufficient mechanical strength. This motivates the development of carefully designed low-$Z$ alloys in which lithium is incorporated to enhance radiation length while maintaining structural stability. Currently, materials such as beryllium, titanium, aluminum alloys, and stainless steel have been employed as beampipe materials owing to their balanced combinations of mechanical strength, radiation length, and vacuum compatibility~\cite{Veness:2002ab, Ayres:1981zz, knaster, Brunet:1979tq}.  
For commonly used stainless steel beampipes, this figure of merit is relatively low ($X_{0}E^{1/3}\approx 0.1017$), reflecting excellent mechanical robustness but poor radiation transparency. Aluminum alloys offer an improvement in radiation length, yielding $X_{0}E^{1/3}\approx 0.37$, and are therefore widely adopted in MeV-GeV energy scale accelerator facilities. For example, in the BM@N experiment at the Nuclotron-based Ion Collider Facility (NICA), a combination of stainless steel, carbon fiber, aluminum, and titanium sections is employed to satisfy region-specific mechanical and vacuum requirements~\cite{Afanasiev:2023opv}. Similarly, synchrotron radiation sources such as Indus-2 in India~\cite{Deb_2013,Bhawalkar1998}, operating at MeV-scale beam energies, utilize aluminum alloy 5083-H321 beampipes with wall thicknesses of approximately 0.2~mm, achieving a figure of merit of $X_{0}E^{1/3}\approx 0.374$~\cite{Singh:2024ihp}. 
While these configurations are adequate for existing facilities, the relatively modest values of $X_{0}E^{1/3}$ highlight the intrinsic limitations of conventional materials. Even a moderate enhancement in this figure of merit can translate into a measurable reduction in multiple scattering and background generation, thereby improving track reconstruction efficiency, momentum resolution, and overall data quality. Consequently, developing beampipe materials with higher $X_{0}E^{1/3}$ is directly relevant for next-generation accelerator experiments. These experiments will require increasingly precise measurements with strict control of material-induced systematic effects.

This study is motivated by the need to improve the figure of merit of conventional beampipe materials.
While materials such as aluminum, titanium, and stainless steel provide reliable mechanical performance and vacuum compatibility, their relatively low radiation lengths further increase the material-induced beam degradation. 
In this context, Mg--Al--Li alloys offer several intrinsic advantages. The incorporation of low-$Z$ elements, such as lithium and magnesium, substantially enhances the radiation length, while aluminum provides structural stability, good vacuum compatibility, and better manufacturability. This combination enables systematic improvement of the figure of merit relative to conventional beampipe materials without sacrificing mechanical integrity. To identify promising compositions and ensure thermodynamic viability, the Thermo-Calc software suite is employed to analyze ternary Mg–Al–Li phase diagrams, phase equilibria, and solidification behavior. This allows candidate alloy compositions to be selected from regions exhibiting stable or technologically relevant phase fields. To quantitatively assess the mechanical performance of selected Mg--Al--Li alloys, we employ first-principles density functional theory calculations using the Vienna \textit{Ab Initio} Simulation Package (VASP)~\cite{Kresse1996} to evaluate elastic properties and estimate mechanical responses based on alloy phase composition. This integrated approach of selecting alloys using a thermodynamic simulation technique for obtaining radiation length, as well as evaluation of mechanical properties by DFT calculation, allows a quantitative assessment of the designed alloys with improved figure of merit for beampipe applications.

Specifically, in this study, two aluminum-rich alloys, A1 ($\mathrm{Al_{61.5}Li_{10.8}Mg_{27.7}}$) and A2 ($\mathrm{Al_{66}Li_{19.4}Mg_{14.6}}$), together with three magnesium-rich alloys, M1 ($\mathrm{Al_{23.9}Li_{29.3}Mg_{46.8}}$), M2 ($\mathrm{Al_{19}Li_{20.6}Mg_{60.4}}$), and M3 ($\mathrm{Al_{39.8}Li_{20.1}Mg_{40.1}}$), are analyzed in terms of thermodynamic phase stability, density, radiation length, elastic modulus, and the combined performance metric $X_0E^{1/3}$.
Although the elastic moduli of the investigated Mg–Al–Li alloys are comparable to those of conventional beampipe materials, their much higher radiation lengths lead to an improved figure of merit. This improvement highlights the advantage of Mg–Al–Li alloys in reducing material-induced beam degradation while maintaining adequate mechanical rigidity. As a result, these alloys emerge as promising candidates for beampipe materials in next-generation accelerators. This motivates the exploration of Mg–Al–Li alloy systems to achieve a higher figure of merit than currently used materials.

The remainder of this paper is organized as follows. Section~\ref{Sec-formalism} outlines the formalism for material property estimations, including radiation length and elastic modulus calculations. The results and their comparison with conventional materials such as stainless steel and aluminum alloys are presented in Section~\ref{Sec-results}. Finally, Section~\ref{sec-summary} summarizes the main conclusions and discusses the outlook for further alloy development in the context of high-energy accelerator technology.

\section{Formalism}
\label{Sec-formalism}

In this section, we discuss the formalism used to calculate the radiation length and the elastic modulus of the proposed alloys. The radiation length of the alloys is evaluated using a mixture rule based on weight fractions of constituent elements~\cite{DaSilva2025}. The elastic modulus is estimated using first-principles calculations based on density functional theory. A brief description of thermodynamic modeling using Thermo-Calc software to perform phase calculations for proposed alloys is also presented here.

\subsection{Calculation of radiation length}

The radiation length is defined as the mean distance over which a high-energy electron loses  $1/e$ of its energy through bremsstrahlung and multiple scattering processes. For an electron or a charged particle with initial energy $\epsilon_0$ traversing a distance $x$ in a material, the energy attenuation follows,
\begin{align}
    \epsilon = \epsilon_0 e^{-x/X_0}.
\end{align}

We employ the following approach to estimate $X_0$ for the alloy systems. For a given element, the radiation length constant $\Bar{X_0}$ (expressed in $kg/m^2$) can be calculated using the semi-empirical relation~\cite{Stolzenberg:2019ldl, PhysRevAccelBeams.27.024801},
\begin{align}
\label{eq:X0_element}
    \Bar{X_0} = \frac{716.4 A}{Z(Z+1)\ln\left(\frac{287}{\sqrt{Z}}\right)} \quad (kg/m^2),
\end{align}
where $Z$ is the atomic number and $A$ is the atomic mass number of the element. To obtain the radiation length $X_0$ in units of meters, the radiation length constant is divided by the material's mass density $\rho$ ($kg/m^3$),
\begin{align}
    X_0 = \frac{\Bar{X_0}}{\rho} \quad (m).
\end{align}
This method provides a quick and analytic estimate of $X_0$ for pure elements.

Further, for multi-component systems such as Mg--Al--Li alloys, a more precise way to estimate $X_0$ is obtained using the weighted inverse sum of the radiation lengths of constituent elements~\cite{DaSilva2025}, given as
\begin{align}
    \frac{1}{X_0} = \sum_{i=1}^{n} \frac{w_i}{X_{0,i}},
\end{align}
where $w_i$ is the weight fraction of the $i$th element and $X_{0,i}$ is the radiation length of the pure element $i$. This method accounts for compositional effects and is recommended for precise modeling of radiation transport in alloy systems.

\subsection{Calculation of elastic modulus}

The elastic modulus of the designed alloys is calculated using density functional theory as implemented in the VASP. The crystallographic input files (CIFs) for all phases considered in this work were obtained from the Materials Project database~\cite{Jain2013MP}. The Perdew-Burke-Ernzerhof (PBE) exchange-correlation functional, within the generalized gradient approximation (GGA), is employed, along with projector augmented-wave (PAW) pseudopotentials. A plane-wave cutoff energy of 520~eV is used, and $k$-point meshes $3\times3\times 3$ are generated using the Monkhorst-Pack scheme with sufficient density to ensure energy convergence. An energy difference of $10^{-4}$~eV was used as the convergence criterion for electronic relaxation. The atomic positions were optimized until the force on each atom was less than 0.01~eV/\AA. Spin polarization was not included in the present calculations. 

 The elastic modulus, bulk modulus ($K$), and shear modulus ($G$) are derived from the full stiffness tensor using the Voigt--Reuss--Hill averaging method. These mechanical properties provide insight into the alloy's structural stability and suitability for integration into accelerator beamlines where high vacuum and mechanical resilience are essential. We evaluated 21 independent elastic constants ($C_{ij}$) and averaged them to obtain the effective cubic elastic parameters: 
$\langle C_{11} \rangle = (C_{11} + C_{22} + C_{33})/3$, 
$\langle C_{12} \rangle = (C_{12} + C_{23} + C_{31})/3$, and 
$\langle C_{44} \rangle = (C_{44} + C_{55} + C_{66})/3$. 

Under the harmonic approximation and within the small-deformation limit, the change in total energy due to strain can be expressed as~\cite{Nishibuchi2023},
\begin{equation}
    \Delta E(V, \{\varepsilon_i\}) = E(V, \{\varepsilon_i\}) - E(V_0, 0) = 
    \frac{V_0}{2} \sum_{i,j=1}^{6} C_{ij} \varepsilon_i \varepsilon_j,
\end{equation}
where $E(V_0, 0)$ and $E(V, \{\varepsilon_i\})$ denote the total energies of the equilibrium and deformed structures with volumes $V_0$ and $V$, respectively. The quantities $\varepsilon_i$ and $\varepsilon_j$ represent the six independent components of the strain vector ($1 \leq i,j \leq 6$), and $C_{ij}$ are the elements of the symmetric $6\times6$ elastic stiffness tensor. The elastic stiffness tensor was computed from the second derivative of the total energy with respect to strain, following the energy–strain relationship~\cite{LePage2001}.

The bulk modulus and shear modulus were derived from the stiffness tensor using the Voigt ($K_V$ and $G_V$)~\cite{Voigt1910} and Reuss ($K_R$ and $G_R$)~\cite{Reuss1930} averaging schemes. The Voigt bounds are given by,
\begin{align}
    9K_V &= (C_{11} + C_{22} + C_{33}) + 2(C_{12} + C_{23} + C_{31}), \label{eq:KV} \\
    15G_V &= (C_{11} + C_{22} + C_{33}) - (C_{12} + C_{23} + C_{31}) + \nn\\
    &~~~~~3(C_{44} + C_{55} + C_{66}). \label{eq:GV}
\end{align}
The corresponding Reuss bounds can be expressed as,
\begin{align}
    \frac{1}{K_R} &= (S_{11} + S_{22} + S_{33}) + 2(S_{12} + S_{23} + S_{31}), \label{eq:KR} \\
    \frac{15}{G_R} &= 4(S_{11} + S_{22} + S_{33}) - 4(S_{12} + S_{23} + S_{31}) +\nn\\&~~~~~ 3(S_{44} + S_{55} + S_{66}), \label{eq:GR}
\end{align}
where $S_{ij}$ are the components of the compliance tensor, defined as the inverse of the stiffness tensor ($[S_{ij}] = [C_{ij}]^{-1}$).

The arithmetic mean of the Voigt and Reuss bounds provides the Voigt–Reuss–Hill (VRH) average, representing the effective isotropic moduli of the polycrystalline material~\cite{Hill1952},
\begin{align}
    K_{\text{VRH}} &= \frac{K_V + K_R}{2}, \\
    G_{\text{VRH}} &= \frac{G_V + G_R}{2}.
\end{align}
The Elastic modulus is then obtained using,
\begin{equation}
    E = \frac{9K_{\text{VRH}}G_{\text{VRH}}}{3K_{\text{VRH}} + G_{\text{VRH}}}.
\end{equation}

\subsection{Phase evolution study using thermodynamic simulation }

We employ the Thermo-Calc software suite to perform equilibrium and non-equilibrium phase calculations for various compositions of Mg--Al--Li alloys. The thermodynamic simulations are carried out using the TCAL8 database for alloys A1 and A2, which is specifically tailored for aluminum-based systems and includes relevant thermodynamic descriptions for binary and ternary phases involving Al, Li, and Mg. The TCMG6 (v6.3) dataset is used for thermodynamic simulations of M1, M2, and M3 alloys, which are specifically tailored for magnesium-based systems.

The simulations are conducted via TC-Python, an integrated scripting environment that enables automated high-throughput exploration of compositional design space. In each simulation, the components and phases of interest are specified, and the equilibrium phase diagrams are studied over a temperature range relevant for alloy solidification and processing.
To more realistically capture the solidification behavior observed in industrially processed alloys, we consider non-equilibrium solidification pathways using the Gulliver--Scheil model~\cite{samal2016solidification, zuo2013evolution}. This model assumes complete mixing in the liquid phase and no diffusion in the solid phase, making it suitable for representing solute segregation during rapid solidification. Under these assumptions, the liquid concentration at the solid--liquid interface is given by:
\begin{align}
   C_L = C_0 (1 - f_s)^{k-1},
\end{align}
where $C_0$ is the initial liquid concentration, $k$ is the equilibrium partition coefficient, and $f_s$ is the solid fraction. The Gulliver--Scheil simulations provide insight into phase formation sequences and solidification routes, which are critical for predicting microstructural evolution of the alloy. The phase formation and solidification behavior of the five designed Mg--Al--Li alloys can also be interpreted using the invariant reactions listed in Table~3 of Ref.~\cite{Wang}.

\section{Results and Discussion}\label{Sec-results}
\begin{figure*}
	\centering
	\includegraphics[width=0.45\linewidth]{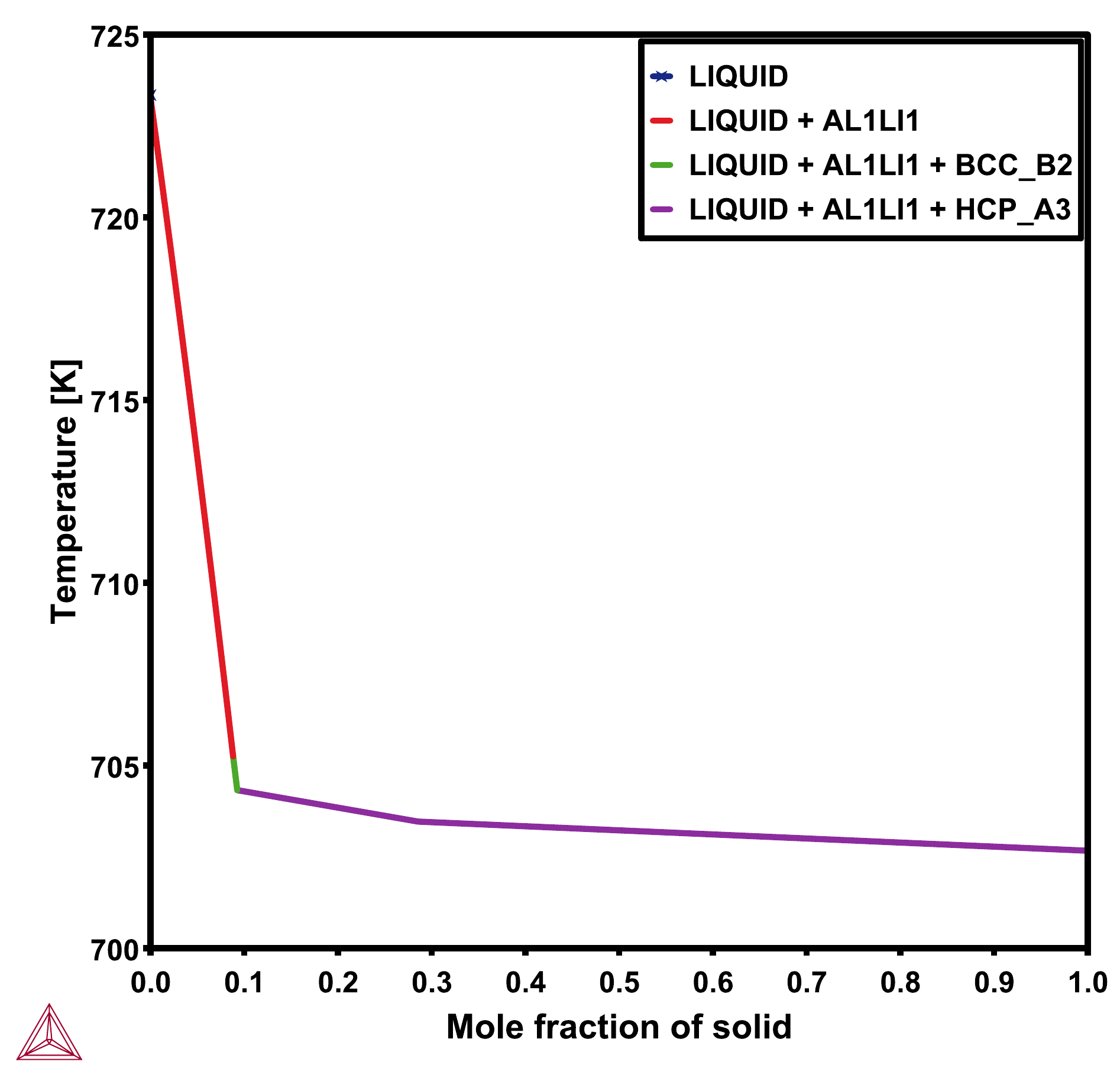}
 \includegraphics[width=0.45\linewidth]{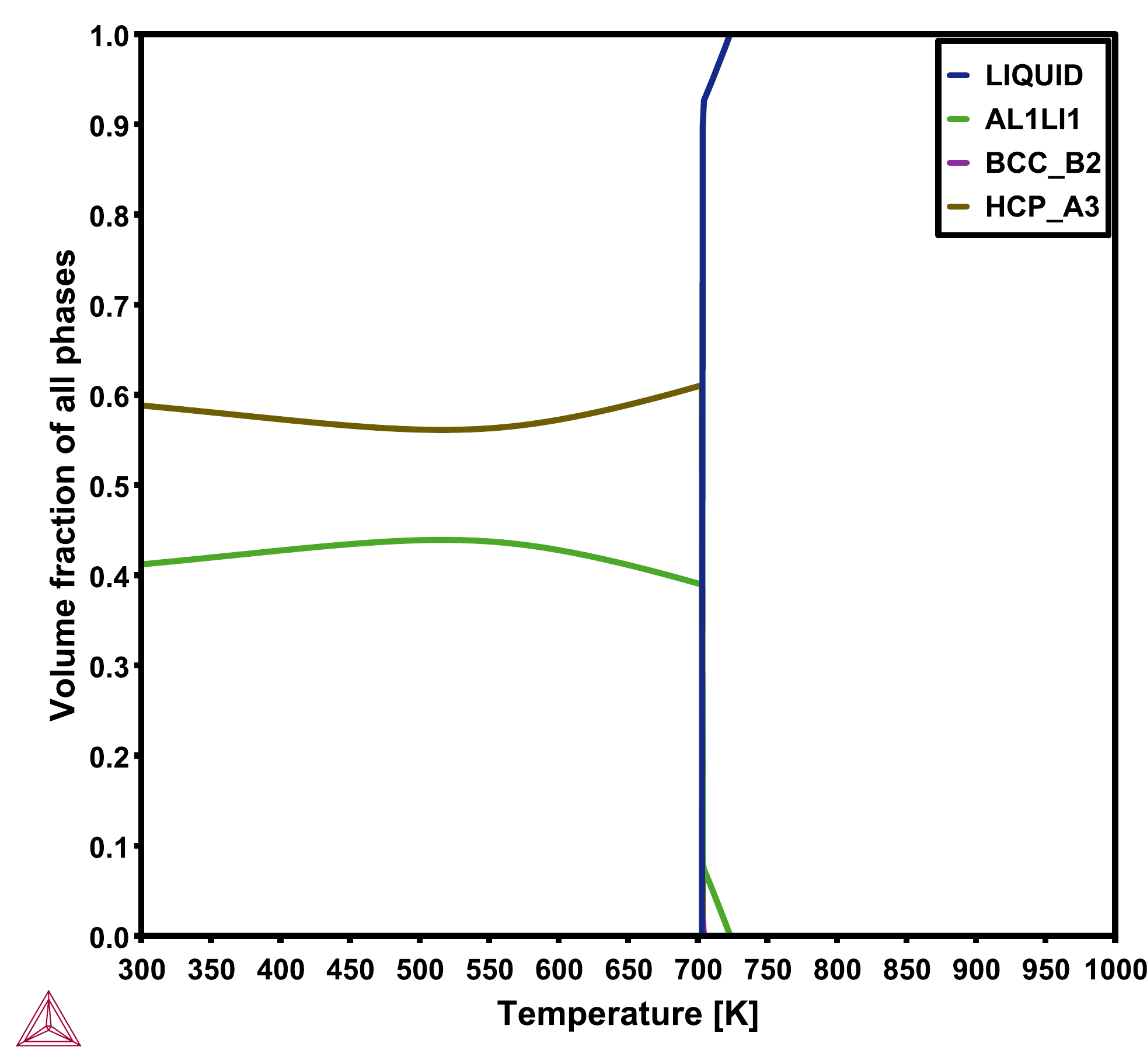}
	\caption{For alloy-M1, (left panel) mole fraction of solids and (right panel) volume fraction of solids.}
	\label{Fig-alloy1}
\end{figure*}
\begin{figure*}
	\centering
	\includegraphics[width=0.45\linewidth]{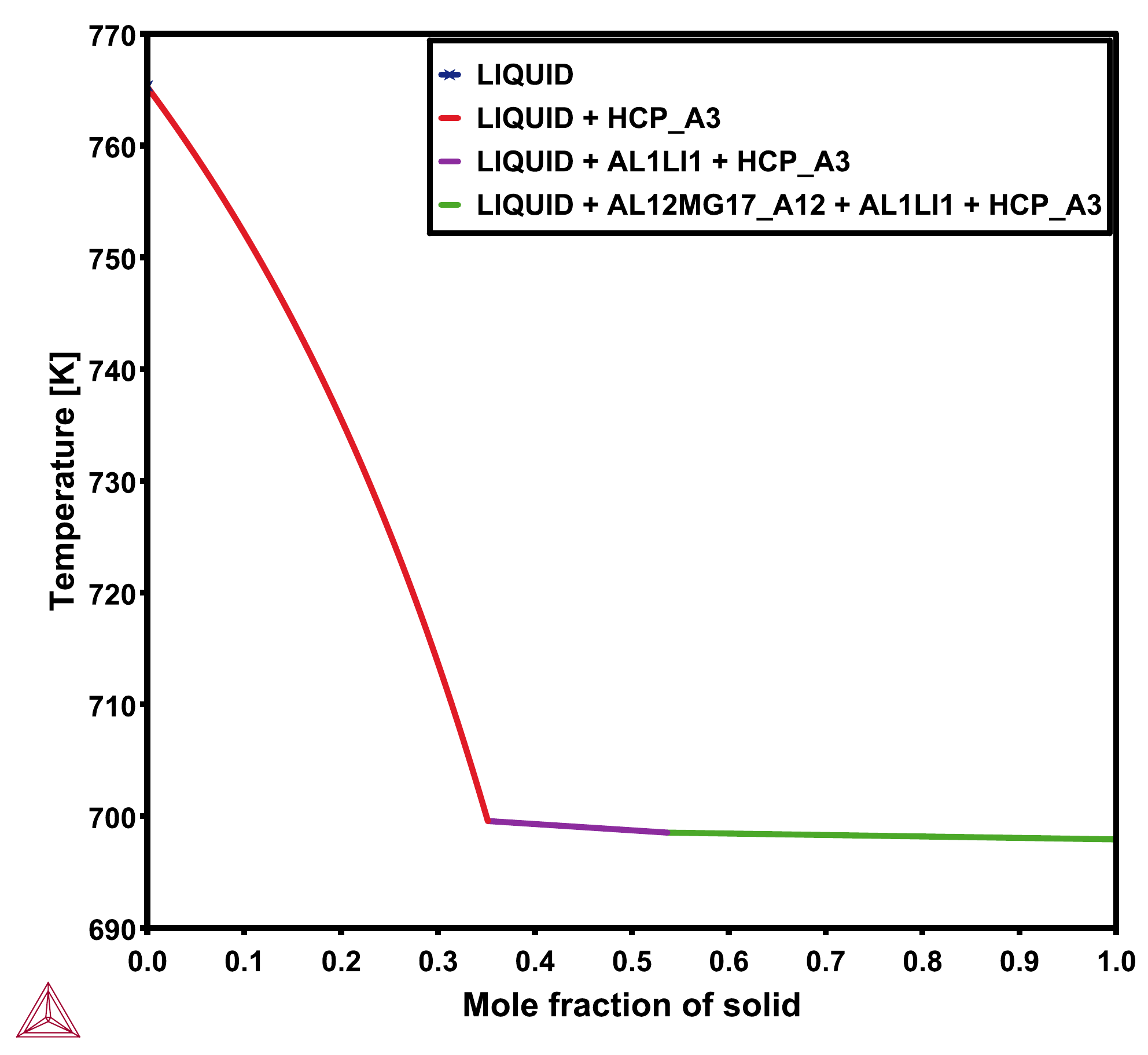}
 \includegraphics[width=0.45\linewidth]{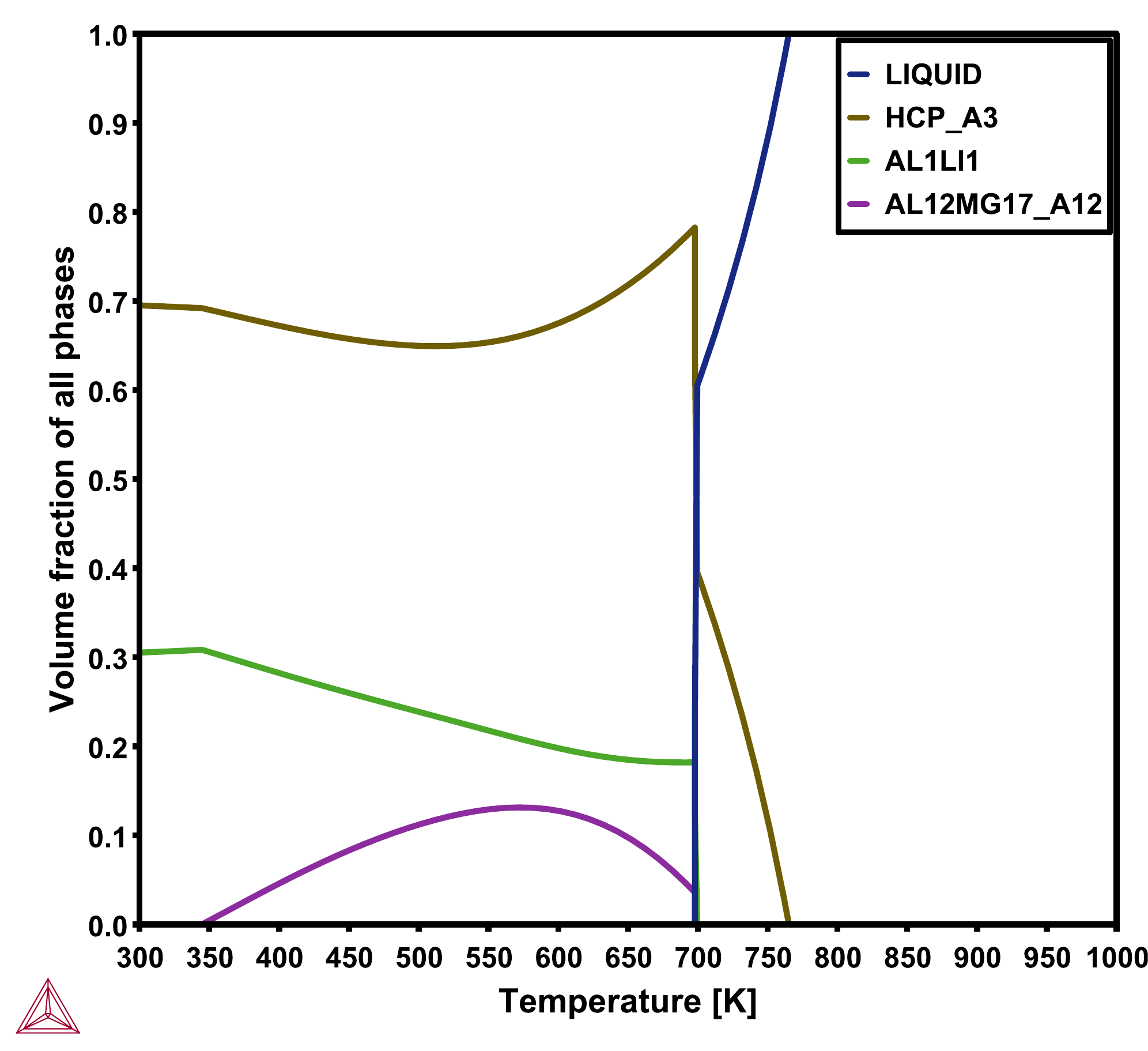}
\caption{For alloy-M2, (left panel) mole fraction of solids and (right panel) volume fraction of solids.}
	\label{Fig-alloy2}
\end{figure*}
\begin{figure*}
	\centering
\includegraphics[width=0.50\linewidth]{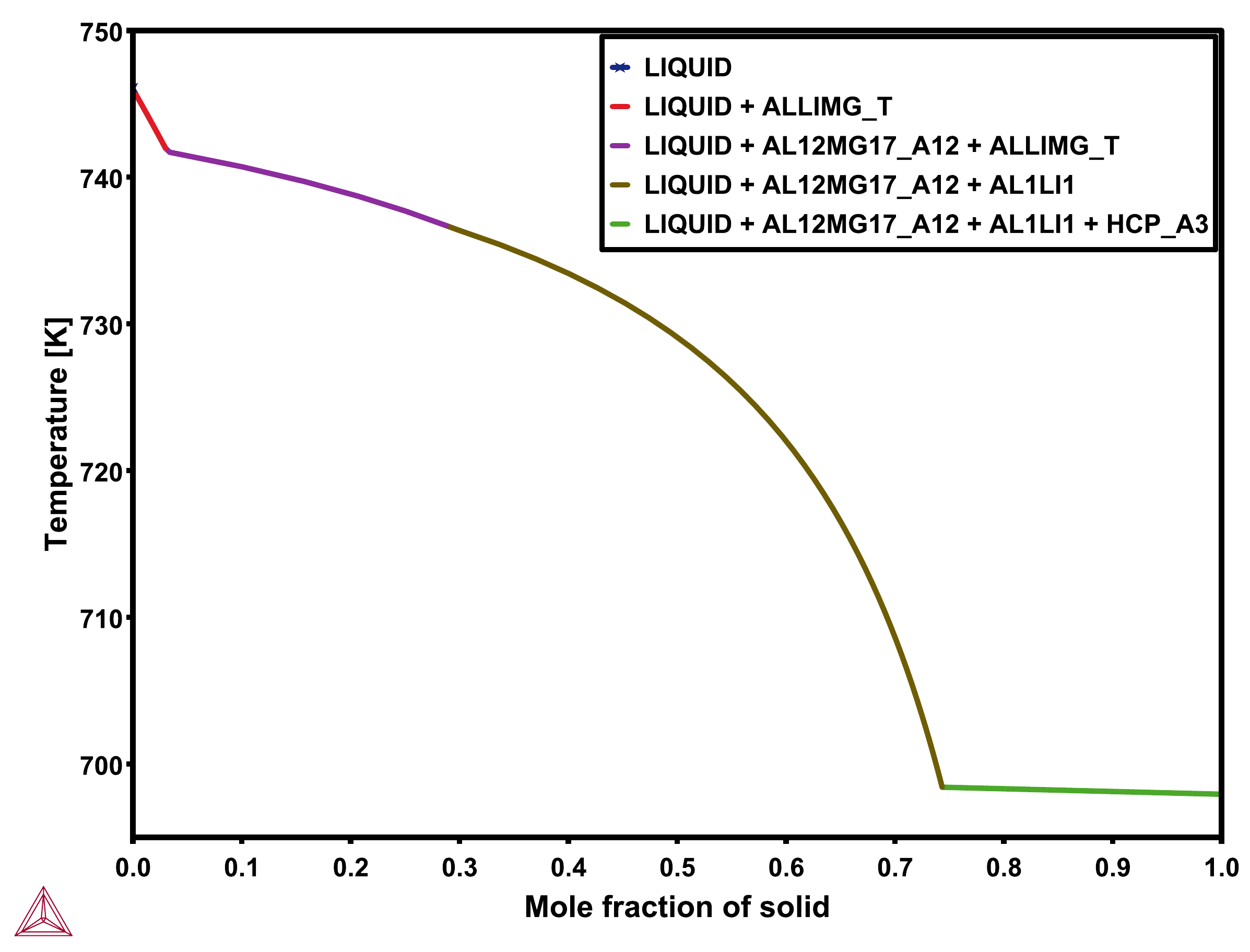}
 \includegraphics[width=0.40\linewidth]{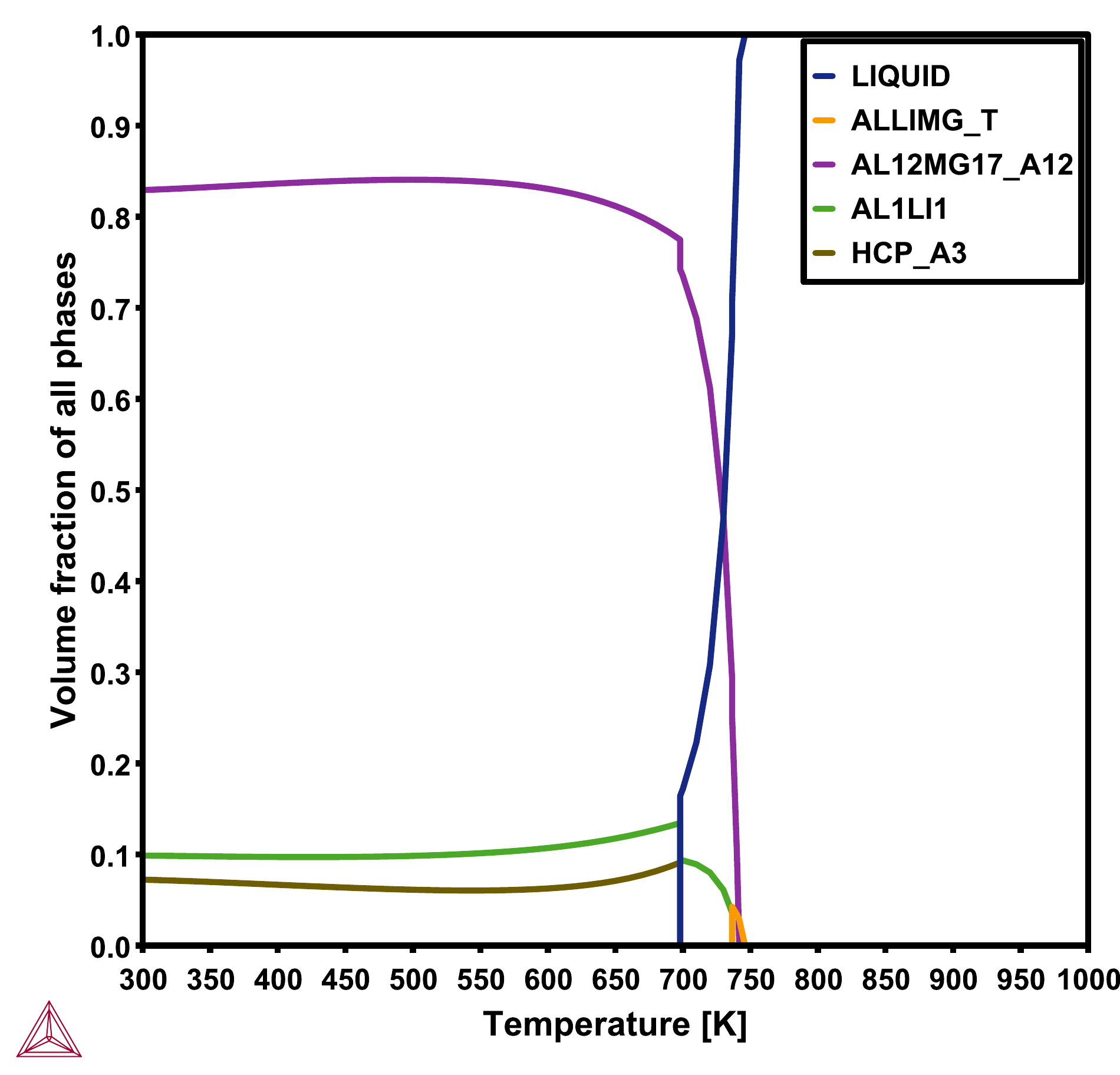}
\caption{For alloy-M3, (left panel) mole fraction of solids and (right panel) volume fraction of solids.}
	\label{Fig-alloy3}
\end{figure*}
\subsection{Solidification behavior of Mg--Al--Li alloys}

The equilibrium solidification behavior of the three magnesium-rich alloys M1, M2, and M3 was investigated using Thermo-Calc with the TCMG6 (v6.3) thermodynamic database. The evolution of solid phases during cooling is quantified in terms of both mole fraction and volume fraction as a function of temperature, as shown in Figs.~\ref{Fig-alloy1}--\ref{Fig-alloy3}. These representations provide complementary insights into phase formation and the resulting microstructural constitution relevant for beampipe applications.

\subsubsection{Alloy M1}

Figure~\ref{Fig-alloy1} presents the solidification path of alloy M1. The left panel shows the mole fraction of solids as a function of temperature, indicating that solidification initiates slightly above $\sim720$~K with the precipitation of the primary Al1Li1 [i.e., (AlLi)] phase from the initial liquid.
\begin{equation}
\mathrm{L} \;\rightarrow\; \mathrm{Al1Li1} + \mathrm{L}.
\end{equation}
As cooling proceeds, additional solid phases appear through successive monovariant reactions involving $\rm {(L, Al1Li1, BCC\_B2 )}$ and $\rm {(L, Al1Li1, HCP\_A3 )}$ phases. 
\begin{equation}
\mathrm{L} \;\rightarrow\;   \mathrm{Al1Li1} + \mathrm{BCC\_B2},
\end{equation}
\begin{equation}
\mathrm{L} \;\rightarrow\;  \mathrm{Al1Li1} +\mathrm{HCP\_A3},
\end{equation}
where BCC\_B2 is (Li)$_{ss}$ and HCP\_A3 is (Mg)$_{ss}$ phases. This leads to a multiphase solid mixture at lower temperatures. Based on the above two monovariant reactions, alloy M1 undergoes an invariant reaction, as
\begin{equation}
\mathrm{L} + \mathrm{BCC\_B2} \;\rightarrow\; \mathrm{HCP\_A3} + \mathrm{AlLi},
\end{equation}
This studied alloy shows a unique phase equilibrium with the formation of multiple stable phases.

The right panel of Fig.~\ref{Fig-alloy1} illustrates the corresponding volume fraction of each phase with respect to temperature during solidification. Near the liquidus, the HCP\_A3 phase dominates the solid fraction, reflecting the Mg-rich nature of alloy M1. With decreasing temperature, the volume fraction of Al1Li1 increases steadily, while the BCC\_B2  phase remains comparatively minor. The coexistence of HCP\_A3 and Al1Li1 over a broad temperature range. This alloy shows a stable multiple-phase mixture, exhibiting low density due to the presence of low-$Z$ lithium-containing phases.

\subsubsection{Alloy M2}

The solidification behavior of alloy M2 is shown in Fig.~\ref{Fig-alloy2}. Compared to M1, alloy M2 exhibits a higher liquidus temperature, reflecting its increased aluminum content. The left panel indicates that solidification begins with the formation of the HCP\_A3 phase,
\begin{equation}
\mathrm{L} \;\rightarrow\; \mathrm{HCP\_A3} + \mathrm{L},
\end{equation}
followed by monovariant reaction involving $\rm {(L, Al1Li1, HCP\_A3 )}$. Finally, formation of a multi-phase mixture via an invariant reaction involving $\rm {(L, Al12Mg17\_A12, Al1Li1, HCP\_A3 )}$. The invariant reaction is given as,
\begin{equation}
\mathrm{L} \;\rightarrow\; \mathrm{HCP\_A3}
+ \mathrm{Al1Li1}
+ \mathrm{Al12Mg17\_A12}.
\end{equation}

The volume fraction evolution shown in the right panel reveals that the HCP\_A3  phase remains the dominant constituent across most of the solidification range. However, the contribution of Al12Mg17\_A12 becomes increasingly significant at lower temperatures, while Al1Li1 contributes a moderate but non-negligible fraction.

\subsubsection{Alloy M3}

Figure~\ref{Fig-alloy3} shows the solidification characteristics of alloy M3, which possesses a comparatively higher lithium content. The left panel indicates a more complex solidification path, with multiple phase fields encountered as the system cools from the liquid state. Primary solidification initiates with the formation of the intermetallic AlLiMg\_T phase.
\begin{equation}
\mathrm{L} \;\rightarrow\; \mathrm{AlLiMg\_T} + \mathrm{L},
\end{equation}
As cooling proceeds, additional solid phases appear through successive monovariant reactions involving $\rm {(L, Al12Mg17\_A12, AlLiMg\_T )}$ and $\rm {(L, Al12Mg17\_A12, Al1Li1 )}$ phases. 
\begin{equation}
\mathrm{L} \;\rightarrow\; \mathrm{Al12Mg17\_A12} + \mathrm{AlLiMg\_T}.
\end{equation}
\begin{equation}
\mathrm{L} \;\rightarrow\; \mathrm{Al12Mg17\_A12} + \mathrm{Al1Li1}.
\end{equation}
Based upon the above two monovariant reaction, we propose the following invariant reaction,
\begin{equation}
\mathrm{L} + \mathrm{AlLiMg\_T} \;\rightarrow\; \mathrm{Al1Li1} + \mathrm{Al12Mg17\_A12}.
\end{equation}
At lower temperatures, the residual liquid undergoes a final invariant eutectic reaction producing additional $\mathrm{HCP\_A3}$ phase with the following invariant reaction,
\begin{equation}
\mathrm{L} \;\rightarrow\; \mathrm{Al12Mg17\_A12}
+ \mathrm{Al1Li1}
+ \mathrm{HCP\_A3}.
\end{equation}

The corresponding volume fraction plot (right panel of Fig.~\ref{Fig-alloy3}) highlights the strong dominance of lithium-containing intermetallics over a wide temperature range. In particular, Al12Mg17\_A12 occupies a substantial volume fraction prior to complete solidification, while Al1Li1 and HCP\_A3 contribute smaller but stable fractions. It is important to note that the formation of multiple intermetallic phases in the studied alloy system is a good candidate for achieving good mechanical strength.

\subsection{Ternary phase equilibria and liquidus projection}
\begin{figure*}
	\centering
	\includegraphics[width=0.48\linewidth]{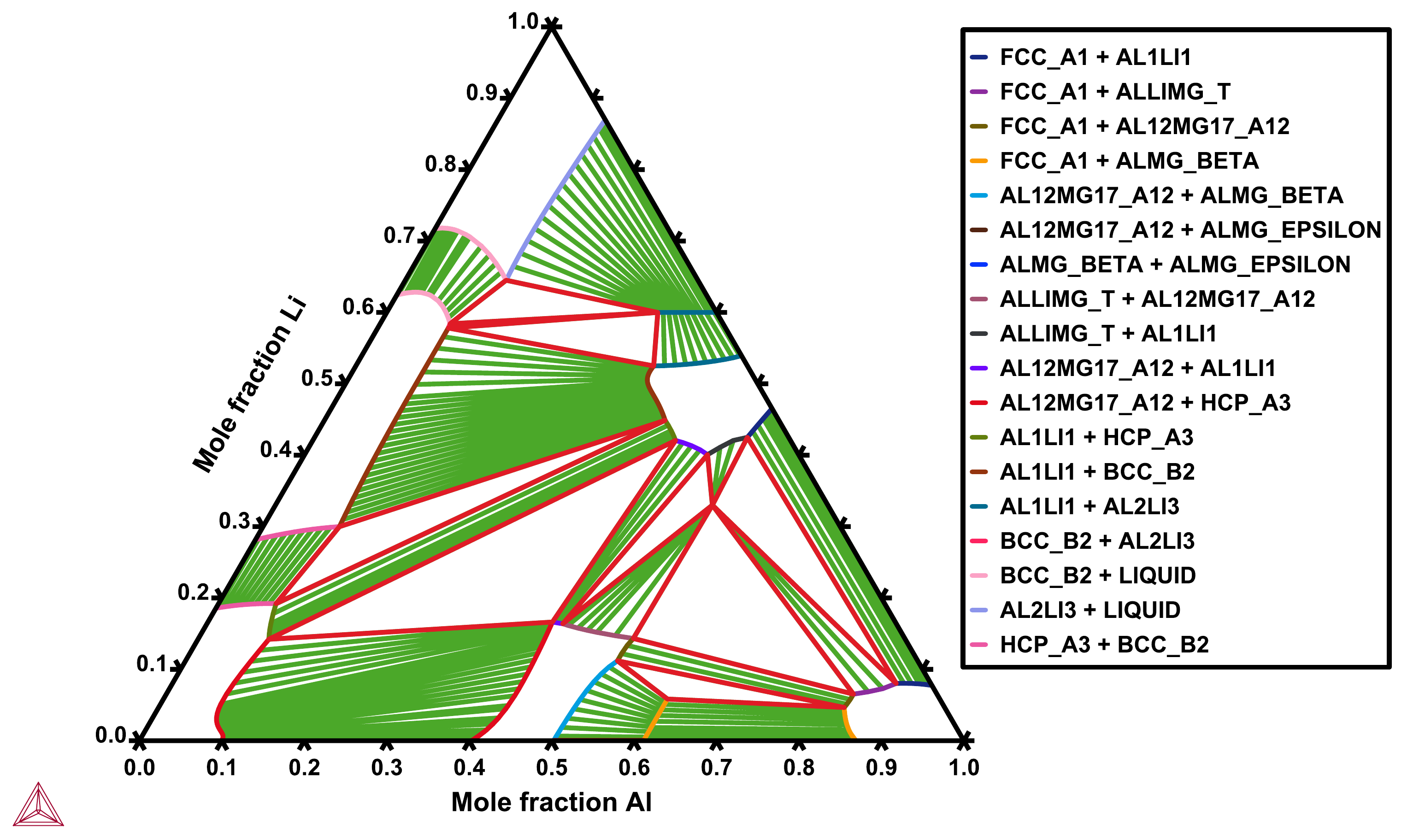}
 \includegraphics[width=0.48\linewidth]{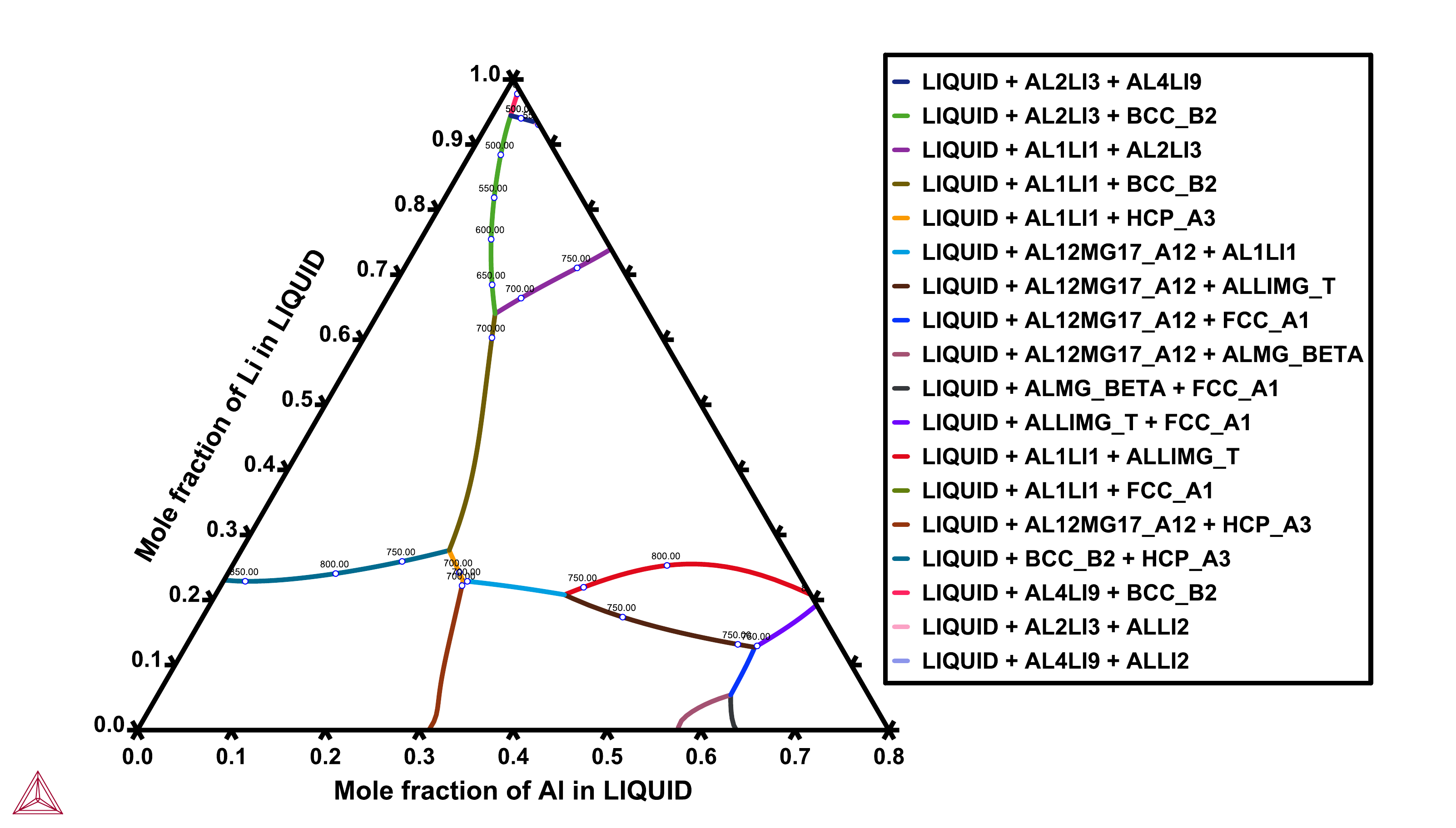}
	\caption{Left: Ternary plot, Right: liquidus projection for AlLiMg based alloy at 673 K.}
	\label{Fig-Ternary}
\end{figure*}
The overall phase stability and solidification trends of  Mg--Al--Li alloys are summarized in Fig.~\ref{Fig-Ternary}. The left panel shows the equilibrium ternary phase diagram at 673 K, revealing extensive multiphase regions involving combinations of HCP\_A3, Al1Li1, and Al12Mg17\_A12. The broad stability fields observed in this diagram underscore the thermodynamic feasibility of tailoring alloy compositions to achieve desired phase fractions.

The right panel of Fig.~\ref{Fig-Ternary} illustrates the monovariant reaction paths in the  Mg--Al--Li ternary system, represented as liquidus projection trajectories on the Gibbs composition triangle. Each curve corresponds to a two-phase equilibrium between the liquid and a single solid phase, evolving continuously with temperature during solidification. The liquidus surface is partitioned into regions corresponding to primary crystallization of HCP\_A3, Al1Li1, Al12Mg17\_A12, and related phases.
Along a monovariant path, the system satisfies the Gibbs phase rule with one degree of freedom, such that the liquid composition changes progressively as two solid precipitates. The direction of each trajectory indicates the evolution of the liquid composition as cooling proceeds, while the associated solid phase remains in equilibrium with the liquid over a finite temperature interval.
The monovariant lines connect primary phase fields to terminal invariant points, resulting in three solid phases in equilibrium with liquid. Transitions between different monovariant paths reflect changes in the stable solid phases as the liquid composition crosses phase boundaries in the ternary diagram.
Such monovariant trajectories provide a thermodynamic map of solidification routes, which is essential for understanding how small variations in alloy composition lead to distinct solidification sequences. The Fig.~\ref{Fig-Ternary}, therefore, serves as a fundamental framework for interpreting phase evolution in multicomponent Mg--Al--Li alloys under near-equilibrium cooling conditions.  The M1, M2, and M3 ternary alloys with selected compositions are consistent with the solidification paths observed in Figs.~\ref{Fig-alloy1}--\ref{Fig-alloy3}, validating the internal consistency of the thermodynamic predictions. 

\subsection{Implications for beampipe applications}

The results demonstrate that Mg--Al--Li alloys can be engineered to exhibit stable multiphase microstructures with tunable phase fractions over a wide temperature range. The coexistence of Mg-rich HCP phases, which provide mechanical support, with lithium-containing intermetallics, which enhance radiation length, is particularly advantageous for accelerator beampipe applications. By appropriate selection of alloy composition, it is possible to optimize the balance between mechanical stiffness, density, and radiation transparency, thereby improving the overall figure of merit relevant for high-precision particle accelerator environments.

\begin{table*}
    \centering
    \begin{tabular}{|c|c|c|c|c|}
    \hline
       Material  & Radiation  & Elastic  & $X_{0}E^{1/3}$ & Density  \\
         & Length $(X_{0})$ [m] & Modulus (E) [GPa] &  & [$\mathrm{kg/m^3}$] \\
    \hline \hline
    \multicolumn{5}{|c|}{\textbf{Al-rich alloys}} \\
    \hline
    A1 
    ($\mathrm{Al_{61.5}Li_{10.8}Mg_{27.7}}$) & 0.1178 & 99.66 & 0.5460 & 2108 \\
    \hline
    A2 ($\mathrm{Al_{66}Li_{19.4}Mg_{14.6}}$) & 0.1240 & 117.43 & 0.6072 & 2035 \\
    \hline
    \hline
    \multicolumn{5}{|c|}{\textbf{Mg-rich alloys}} \\
    \hline
    M1 
    ($\mathrm{Al_{23.9}Li_{29.3}Mg_{46.8}}$) & 0.1703 & 27.538 & 0.5140 & 1559 \\
    \hline
    M2 
    ($\mathrm{Al_{19}Li_{20.6}Mg_{60.4}}$) & 0.1592 & 66.07 & 0.6436 & 1619 \\
    \hline
     M3 ($\mathrm{Al_{39.8}Li_{20.1}Mg_{40.1}}$) & 0.1428 & 104.64 & 0.6729 & 1795 \\
    \hline
     \hline
    \multicolumn{5}{|c|}{\textbf{Benchmark and reference materials}} \\
    \hline
    Stainless Steel 304 & 0.0176 & 193 & 0.1017 & 7930 \\
    \hline
     Aluminum (Al) & 0.089 & 70 & 0.37 & 2699 \\
    \hline
    Al--Ti--V alloy (Ref.~\cite{Singh:2024ihp}) & 0.0756 & 166.2 & 0.416 & 3259 \\
    \hline
    \end{tabular}
    \caption{Comparison of radiation length, elastic modulus, figure of merit, and density for the designed Mg--Al--Li alloys and benchmark beampipe materials.}
    \label{tab:comp}
\end{table*}
\begin{figure*}
    \centering
    \includegraphics[width=0.45\linewidth]{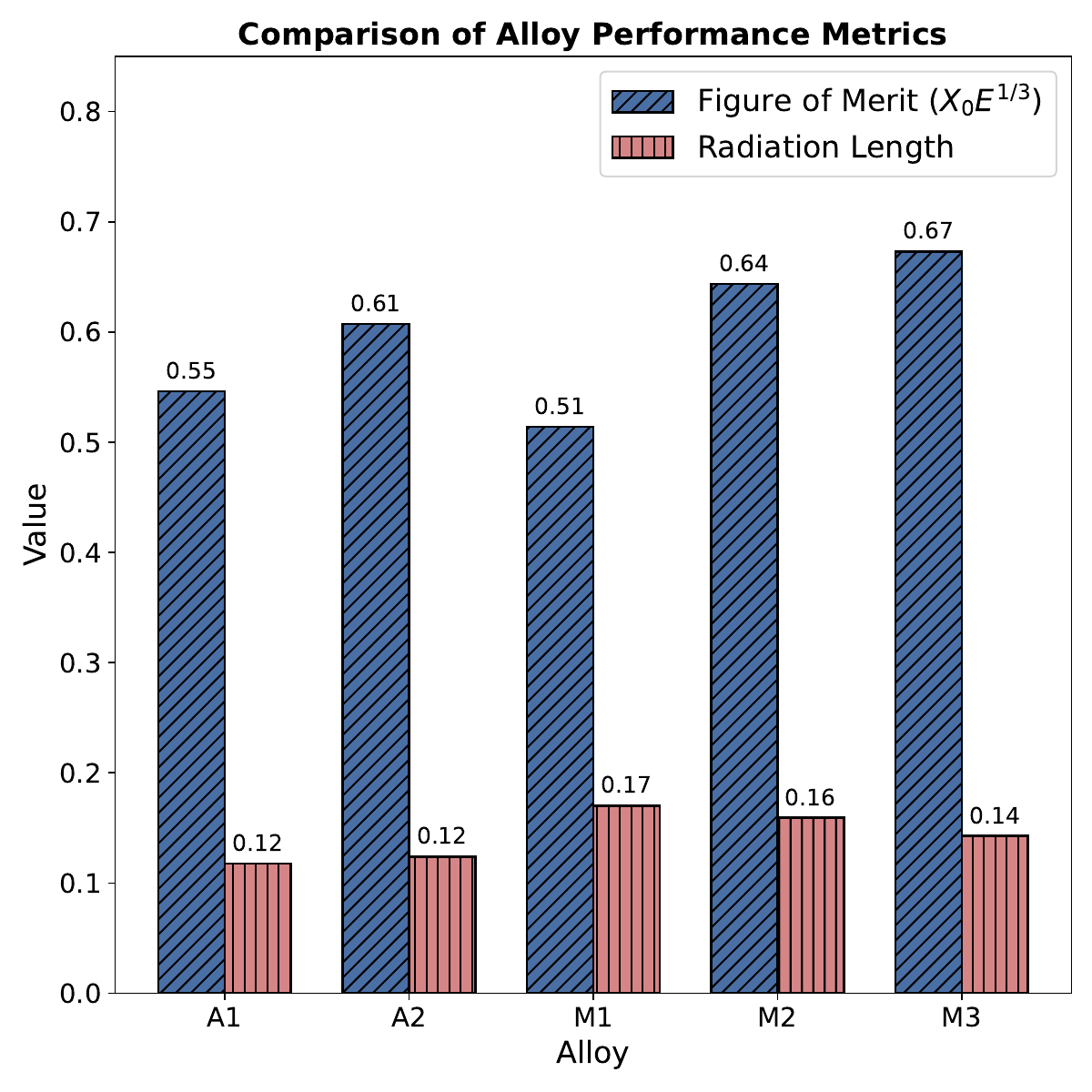}
    \includegraphics[width=0.45\linewidth]{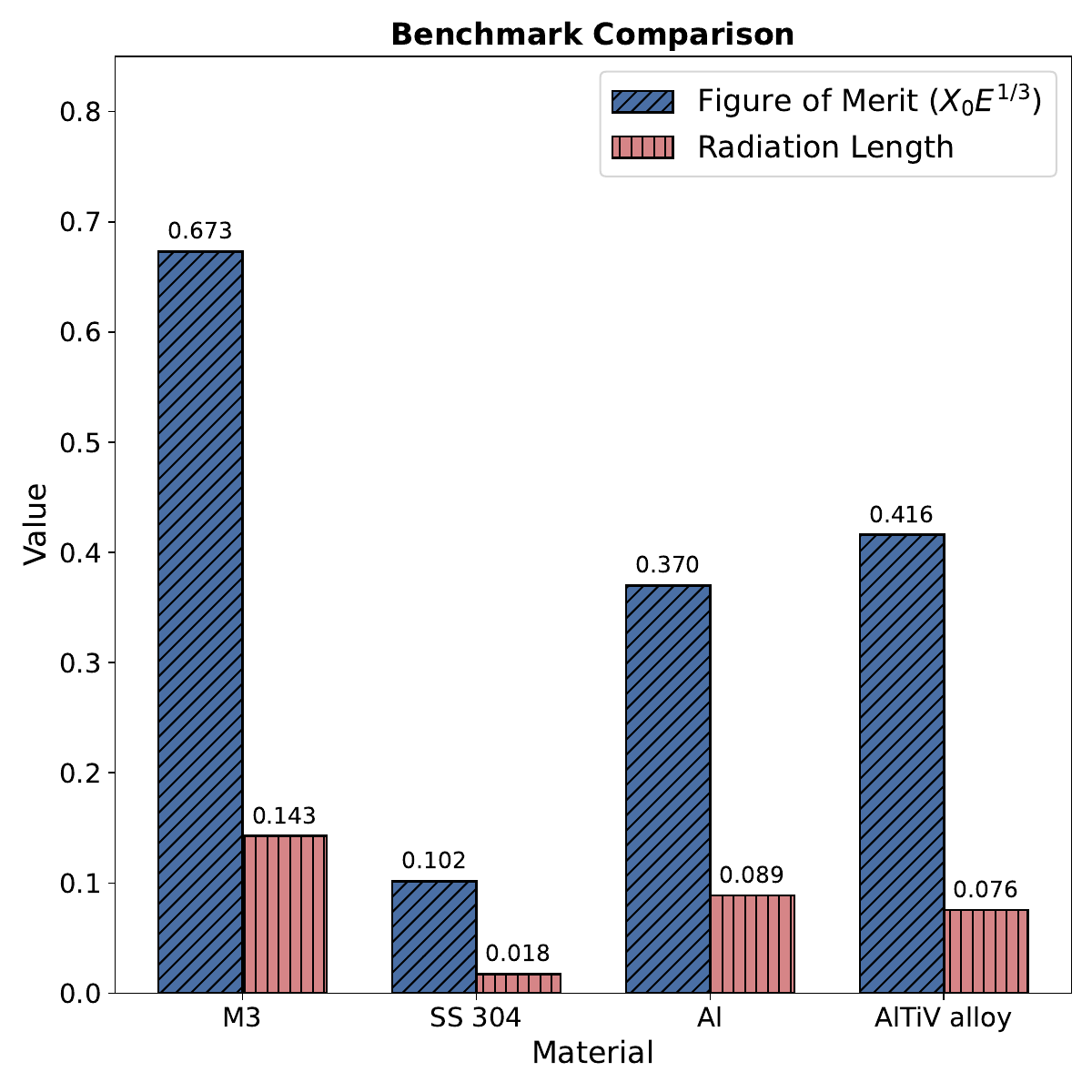}
    \caption{Left: A comparison of lightweight alloys (A1, A2, M1, M2, and M3) in terms of radiation length and figure of merit. Right: Benchmark comparison of M3 alloy with Stainless Steel 304, Aluminum, and Aluminum-rich alloy.}
    \label{fig:bardiag}
\end{figure*}
Table~I presents a comparative summary of radiation length $X_0$, elastic modulus $E$, figure of merit $X_0E^{1/3}$, density, and melting temperature for the designed Mg--Al--Li alloys together with commonly used benchmark beampipe materials. The results clearly demonstrate that targeted compositional tuning within the Mg--Al--Li system leads to a substantial enhancement in radiation length while maintaining competitive mechanical performance.
Among the Al-rich compositions, alloys A1 and A2 exhibit radiation lengths of $X_0=0.1178$~m and $0.1240$~m, respectively, both exceeding that of pure aluminum ($X_0=0.089$~m). This improvement arises from the incorporation of low-$Z$ lithium and magnesium, which effectively reduces the effective atomic number and density of the alloy. The elastic moduli of A1 and A2 remain in the moderate range of $E\approx 100$--120~GPa, resulting in figures of merit $X_0E^{1/3}=0.546$ and $0.607$, respectively. These values represent a clear improvement over conventional aluminum alloys and are significantly higher than those of stainless steel.
A more pronounced enhancement in radiation length is observed for the Mg-rich alloys. Alloy M1 exhibits $X_0=0.1703$~m, nearly twice that of aluminum, albeit with a relatively low elastic modulus of $E=27.5$~GPa. In contrast, alloys M2 and M3 provide a more favorable balance between radiation transparency and mechanical stiffness. Alloy M2 combines a high radiation length of $X_0=0.1592$~m with an elastic modulus of $E=66.1$~GPa, yielding a figure of merit of $X_0E^{1/3}=0.644$. Alloy M3 further improves this balance, achieving $X_0=0.1428$~m and $E=104.6$~GPa, corresponding to the highest figure of merit among all investigated compositions, $X_0E^{1/3}=0.673$.

For comparison, stainless steel 304, despite its high elastic modulus ($E=193$~GPa), exhibits a very low radiation length ($X_0=0.0176$~m), resulting in a small figure of merit of $X_0E^{1/3}=0.102$. Pure aluminum shows a moderate improvement ($X_0E^{1/3}=0.37$), while the Al--Ti--V alloy offers increased stiffness but remains limited by its radiation transparency, yielding $X_0E^{1/3}=0.416$. In contrast, all designed Mg--Al--Li alloys outperform these conventional materials, with figures of merit that are approximately 1.5--2 times higher than aluminum and Al--Ti--V alloys, and up to 5--6 times higher than stainless steel.
Overall, the results establish that Mg--Al--Li alloys constitute a promising materials platform for accelerator beampipe applications. By exploiting the low atomic numbers of lithium and magnesium while retaining aluminum as a stabilizing matrix element, these alloys achieve a significantly improved balance between radiation transparency, density, and mechanical performance.

Figure~\ref{fig:bardiag} provides a visual comparison of the radiation length and figure of merit for the designed Mg--Al--Li alloys and selected benchmark materials. The left panel highlights the systematic variation of these metrics across the alloy series A1, A2, M1, M2, and M3. While the radiation length increases significantly with higher Mg and Li content, the figure of merit reflects the competing influence of elastic stiffness. In particular, alloy M3 exhibits the highest $X_{0}E^{1/3}$ among all investigated compositions, confirming that it offers the most favorable balance between radiation transparency and mechanical performance within the ternary system.
The right panel places the best-performing alloy (M3) in direct comparison with conventional beampipe materials. Despite stainless steel and Al--Ti--V alloys possessing higher elastic moduli, their limited radiation lengths result in substantially lower figures of merit. Pure aluminum performs better than stainless steel but remains clearly inferior to the optimized Mg--Al--Li alloy. This comparison highlights that improvements in radiation length significantly enhance the overall performance metric relevant for beampipe applications, and that appropriately designed low-$Z$ ternary alloys can outperform conventional materials by a substantial margin.

\section{Summary}\label{sec-summary}

In this work, we have systematically investigated lightweight ternary Mg--Al--Li alloys as candidate materials for beampipe applications in particle accelerators. The aim is to improve the figure of merit beyond that of conventional materials. We analyze a set of aluminum-rich and magnesium-rich compositions in terms of radiation length, elastic modulus, density, and the figure of merit $X_{0}E^{1/3}$, which provides a quantitative measure of the trade-off between radiation transparency and mechanical stiffness.

The results, summarized in Table~\ref{tab:comp}, demonstrate that all designed Mg--Al--Li alloys exhibit significantly enhanced radiation lengths compared to commonly used beampipe materials such as aluminum alloys, stainless steel, and Al--Ti--V-based alloys. This enhancement arises primarily from the incorporation of low-$Z$ elements, lithium and magnesium, which effectively reduce the average atomic number while maintaining structural stability through aluminum. Although the elastic moduli of the ternary alloys span a wide range, depending on composition. The optimized balance between stiffness and radiation length leads to substantial improvements in the figure of merit. In particular, the best-performing composition (M3) achieves $X_{0}E^{1/3}\approx 0.67$, corresponding to an improvement of nearly a factor of two over aluminum, about $1.5$ times over Al--Ti--V-based alloys, and approximately 5--6 times over stainless steel.
These improvements have direct implications for accelerator performance. A higher figure of merit translates into reduced multiple scattering and energy loss of charged particles traversing the beampipe wall, which in turn enhances tracking accuracy, vertex resolution, and overall data quality in high-precision experiments. The low densities of the Mg--Al--Li alloys further contribute to minimizing the material budget, making them particularly attractive for regions close to the interaction point where material-induced effects are most critical.
The relevance of these material developments is crucial in the context of future heavy-flavor measurements in the ALICE 3 and other LHC physics programs, where precise vertex reconstruction is one of the most critical experimental requirements. The performance of such measurements is critically dependent on minimizing the material-induced noise in the innermost detector regions. An efficiently designed material for the beampipe with high radiation transparency and mechanical strength, therefore, plays a key role in enabling high-precision heavy-flavor studies, including the reconstruction of short-lived charm and beauty hadrons~\cite{ALICE:2022wwr, Das:2021igk}. In this regard, the proposed Mg–Al–Li alloys constitute a promising starting point; nevertheless, further development and optimization will be required to fully satisfy the stringent performance requirements of next-generation detectors and physics programs. Looking ahead, the selection of materials for beampipe will continue to be driven by the increasingly stringent requirements of high-energy and high-luminosity accelerator facilities.

In the longer term, these developments also open pathways toward addressing more demanding material challenges. Although beryllium continues to serve as a benchmark material owing to its exceptional radiation transparency, its well-known drawbacks, including toxicity, brittleness, high cost, and fabrication and joining difficulties, motivate sustained efforts toward identifying safer and more manufacturable alternatives. Future work will therefore extend the present investigation to evaluate the feasibility of light-weighted multicomponent alloys in regimes traditionally dominated by beryllium, with an emphasis on ultra-high-vacuum compatibility and long-term stability under irradiation. Such studies will be essential for assessing the potential of these alloys as next-generation beampipe materials in advanced accelerator facilities.

\section*{Acknowledgments}
 K.S. gratefully acknowledges the financial aid from UGC, Government of India. K.G. acknowledges the financial support from the Prime Minister's Research Fellowship (PMRF), Government of India. The authors acknowledge the DAE-DST, Government of India funding under the mega-science project “Indian participation in the ALICE experiment at CERN” bearing Project No. SR/MF/PS-02/2021-IITI(E-37123). The authors appreciate the use of the computing facility at IIT Indore.

\end{document}